\documentclass[preprint]{aastex} 
\usepackage{epsfig}
\shorttitle{Possible Strong Gravitational Wave Sources}
\shortauthors{de Araujo, Miranda \& Aguiar}

\begin{document}

\title{Possible Strong Gravitational Wave Sources for the {\it
LISA} Antenna}

\author{Jos\'e C.N. de Araujo, Oswaldo D. Miranda\altaffilmark{1}
and Odylio D. Aguiar}

\affil{Divis\~ao de Astrof\'\i sica - Instituto Nacional de Pesquisas
Espaciais, Av. dos Astronautas 1758, S\~ao Jos\'e dos Campos, 12227-010
SP, Brazil}
\email{jcarlos@das.inpe.br, oswaldo@hbar.wustl.edu, odylio@das.inpe.br}

\altaffiltext{1}{Present address: Department of Physics,
Washington University, Campus Box 1105, One Brookings Drive, St.
Louis - MO 63130-4899 - USA.}

\begin{abstract}
Recently Fuller \& Shi proposed that the gravitational collapse of
supermassive objects ($M \gtrsim 10^4M_\odot$) could be a
cosmological source of $\gamma$-ray bursts (GRBs). The major
advantage of their model is that supermassive object collapses are
far more energetic than solar mass-scale compact mergers. Also, in
their proposal the seeds of supermassive black holes (SMBHs) thus
formed could give rise to the SMBHs observed at the center of many
galaxies. We argue here that, besides the generation of GRBs,
there could well occur a strong generation of gravitational waves
(GWs) during the formation of SMBHs. As a result, the rate of such
GW bursts could be as high as the rate of GRBs in the model by
Fuller \& Shi. In this case, the detection of GRBs and bursts of
GWs should occur with a small time difference. We also argue that
the GWs produced by the SMBHs studied here could be detected when
the Laser Interferometric Space Antenna ({\it LISA}) becomes
operative.
\end{abstract}

\keywords{gravitational waves -- supermassive black holes}

\section{Introduction}

The Laser Interferometric Space Antenna ({\it LISA}) is designed
to detect low frequency gravitational waves in the frequency range
$10^{-4} - 1\,\, {\rm Hz}$, which cannot be detected on the Earth
because of seismic noise. A lot of very interesting astrophysical
phenomena are believed to generate GWs in this frequency band: the
formation of supermassive black holes (SMBHs), SMBH-SMBH binary
coalescence, compact stars orbiting SMBHs in galactic nuclei,
pairs of close white dwarfs, pairs of neutron stars, neutron star
and black hole binaries, pairs of contact normal stars, normal
star and white dwarf binaries, and pairs of stellar black holes.

We are particularly concerned here with SMBHs, which are believed
to be present in galactic nuclei (Blandford 1999). Lynden-Bell
(1969) originally proposed that active galaxies harbor a SMBH
engine fed by accretion and there is now solid observational evidence
for this (Richstone {\it et al.} 1998), although there remain some
unanswered questions related to their formation. Several
interesting papers study the mass function of SMBHs in galaxies
(Franceschini, Vercellone \& Fabian 1998; Salucci {\it et al.} 1999),
using different combinations of optical, infrared, radio and X-ray data.

SMBHs could form through the dynamical evolution of dense star
cluster objects; by the merging of SMBHs of smaller masses and by
the viscous evolution and collapse of self-gravitating gaseous
objects (e.g., supermassive stars). Quinlan \& Shapiro (1990)
assumed the existence of a dense star cluster in a galactic
nucleus and followed the build-up of $100 M_\odot$ or larger seed
black holes by collisions. Another possibility is that $\sim 10^6
M_\odot$ SMBHs form by coherent collapse in galactic nuclei before
most of the bulge gas turns into stars (Silk \& Rees 1998;
Haehnelt, Natarajan \& Rees 1998). Other interesting studies
concerning SMBH formation are discussed by Rees (1997, 1998);
Haehnelt \& Rees (1993); Haehnelt (1994); Eisenstein \& Loeb
(1995); Umemura {\it et al.} (1993) and Fuller \& Shi (1998;
hearafter FS).

SMBHs may produce a strong GW signal during their formation, which
could be detectable by {\it LISA} even at cosmological distances.
Since most galaxies could harbor SMBHs it is argued that the
number of events expected could be several per year or even per
day.

It is worth studying whether other astrophysical phenomena related
to the formation of such putative SMBHs, such as the emission of
electromagnetic radiation and neutrinos, could help constrain the
SMBH production rate and formation epoch. For example,
$\gamma$-ray could be related to the production of GWs since the
formation of SMBHs may be a very energetic phenomenon. In
particular GBRs have been puzzling astrophysicists, because of the
enormous electromagnetic energy produced, $\sim
10^{51}-10^{52}\,\, {\rm ergs}$, the spatial isotropy (which
suggests that the sources are cosmological), and the event rate of
several sources per day.

Recently FS (see also Shi \& Fuller 1998; Abazajian, Fuller \& shi
1999) proposed that the gravitational collapse of supermassive
objects ($M \gtrsim 10^4M_\odot$), either as relativistic star
clusters or as a single supermassive star could account for
cosmological GRBs. These authors also proposed that such
supermassive objects should produce neutrino emission, but they did
not consider whether such $\gamma$-ray and neutrino sources could
be also strong GW sources. Since the FS model involves the
formation of a SMBH it is hard to avoid GWs being also produced.

The paper is organized as follows: \S 2 deals with the GWs
generated by GRB SMBHs and  \S 3 presents the discussion and
conclusions.

\section{Gravitational Waves from GRB SMBHs}

This paper extends the study by FS, which considers whether the
collapse of supermassive objects could account for cosmological
GRBs. We argue that such a source of $\gamma$-rays could also be a
strong source of GWs. Then we propose an independent way to check
FS model through GW astronomy.

FS define a supermassive object in terms of a star or star cluster
that undergoes the general relativistic Feynman-Chandrasekhar
instability during its evolution. Supermassive objects with $M
\gtrsim 5\times 10^4 M_\odot$ could leave black hole remnants of
$M \gtrsim 10^3 M_\odot$. To account for the observed rate of GRBs
the supermassive object collapses should  amount to several per
day. Each collapse probably leads to a black hole remnant, so it
is hard to avoid the conclusion that GWs are generated with the
same frequency. If other processes of SMBH formation do not involve GRB
events, GW production rate could well be even higher.

If all supermassive objects form and collapse at a redshift $z$,
as assumed by FS, the event rate is

\begin{equation}
R_{\rm BH} \simeq 4\pi r^2 a_{\rm z}^3 {dr\over dt_0}{\rho_{\rm b}
F(1+z)^3\over M},
\end{equation}

\noindent where $r$ is the Friedman-Robertson-Walker comoving
coordinate of the supermassive object, $a_{\rm z}$  is scale
factor of the Universe at redshift $z$, $t_0$ the age of the
Universe, $\rho_{\rm b}$ is the present value of the baryonic
density, $F$ is the fraction of baryons incorporated in
supermassive objects, $M$ is the mass of the initial hydrostatic
supermassive star, taken to be $M=10M_{\rm BH}$, where
$M_{\rm BH}$ is the mass of the resulting SMBH (FS; Shi \& Fuller
1998; Abazajian, Fuller \& Shi 1999). This rate can be rewritten as

\begin{equation}
R_{\rm BH} \simeq 4\pi r^2c\, n_{\rm BH},
\end{equation}

\noindent where $n_{\rm BH}$ is the number density of SMBHs, given
by

\begin{equation}
n_{\rm BH}={\rho_{\rm b} F \over M} .
\end{equation}

\noindent Equation (2) is implicit in the equations derived by
Carr (1980) in a study concerning the generation of GWs from
SMBHs.

The GW amplitude associated with the formation of each SMBH is
(Thorne 1987)

\begin{equation}
h_{\rm BH}=\bigg({15\over 2\pi}\varepsilon\bigg)^{1/2}{G\over
c^{2}}{M_{\rm BH}\over r_{0}}
\simeq 7.4\times 10^{-20}\varepsilon^{1/2}\bigg({M_{\rm BH}\over
M_{\odot}}\bigg) \bigg({r_{0}\over 1{\rm Mpc}}\bigg)^{-1},
\end{equation}

\noindent where $\varepsilon$ is the efficiency of generation of
GWs. The collapse to a black hole produces a signal with frequency

\begin{equation}
\nu_{\rm obs} = {1\over 5\pi M_{\rm BH}}{c^{3}\over G}(1+z)^{-1}
\simeq 1.3\times 10^{4}{\rm Hz}\bigg({M_{\odot}\over M_{\rm
BH}}\bigg)(1+z)^{-1}.
\end{equation}

\noindent The ensemble of SMBHs formed should produce a background
of GWs with amplitude

\begin{equation}
h_{\rm BG}^{2} = {1 \over \nu_{\rm obs}}\int h_{\rm BH}^{2} dR_{\rm BH}
\end{equation}

\noindent (de Araujo, Miranda \& Aguiar 2000; Miranda, de Araujo
\& Aguiar 2000), where $dR_{\rm BH}$ is the differential SMBH
formation rate. If the SMBHs are assumed to have the same mass and
formation redshift, as in the FS model, we have

\begin{equation}
h_{\rm BG} = \bigg( {4\pi R^2c\, n_{\rm BH} \over \nu_{\rm obs}}\bigg)^{1/2}
h_{\rm BH}.
\end{equation}

\noindent This equation can be  written as

\begin{equation}
h_{\rm BG}=\bigg({\tau \over \Delta t}\bigg)_0^{1/2} h_{\rm BH},
\end{equation}

\noindent (cf. Carr 1980) where the subscript zero indicates a
present day value, $\tau_0$ is the duration of each burst and
$\Delta t_0$ is the interval between bursts. Unlike Carr, we
assume that the above equation holds only for $(\tau/\Delta t)_0
\gtrsim 1$. These time scales are

\begin{equation}
\tau_0 \simeq {1 \over \nu_{\rm obs},}
\end{equation}

\noindent and

\begin{equation}
\Delta t_0 \simeq {1 \over R_{\rm BH}}.
\end{equation}

\noindent The ratio

\begin{equation}
\bigg({\tau \over \Delta t}\bigg)_0 \simeq {4\pi R^2c\, n_{\rm BH} \over
\nu_{\rm obs}},
\end{equation}

\noindent is called duty-cycle and can be interpreted as the
number of overlapping bursts.

If the bursts overlap, $(\tau/\Delta t)_0$ is greater than 1 and
thus $h_{\rm BG}>h_{\rm BH}$; on the other hand, if $(\tau/\Delta
t)_0$ is less than 1, they do not overlap  and the GW background
is not continuous, but consists of a sequence of spaced bursts
with a mean separation $\sim \Delta t_0$ (see Ferrari, Matarrese \&
Schneider 1999, who consider the case where a non-continuous background
also appears).

The cosmological  model  considered  here has a density parameter
$\Omega_0 = \Omega_{\rm b}=0.1$ and Hubble constant $H_0=50\,{\rm
km}\,{\rm s}^{-1}{\rm Mpc}^{-1}$. For a SMBH formed at redshift
$z\simeq 3$ with mass $10^7M_\odot$, the GWs would be detected at
frequency $\nu_{\rm obs}\simeq 3.3\times 10^{-4}{\rm Hz}$, so the
characteristic duration of the burst is $\tau_0 \simeq 3\times
10^3\;{\rm s}$. If $\Delta t_0 \simeq 1/R_{\rm BH} = 1$
day$^{-1}$, as observed for GRBs, we obtain $4.0\times 10^{-2}$
for the duty-cycle. In this case, a population of SMBHs formed at
$z\simeq 3$ with mass $10^7M_\odot$ cannot produce a background
and one will observe a burst a day with duration $\tau_0$,
amplitude $h_{\rm BH}$ and frequency $\nu_{\rm obs}$.

The results are summarized in Fig. 1 which shows the duty-cycle
($\tau_0/\Delta t_0$) as a function of the mass of the SMBHs, for
the formation redshift range $z=1-5$. We also present, for
comparison, results for $R_{\rm BH}\sim 10$  day$^{-1}$.

The energy density of the GWs can be written in units of the
critical density as

\begin{equation}
\Omega_{\rm GW} ={1\over \rho_{\rm c}} {d\rho_{\rm GW}\over
d\log \nu_{\rm obs}},
\end{equation}

\noindent where $\rho_{\rm c}=3H^{2}/8\pi G$. Equivalently

\begin{equation}
\Omega_{\rm GW} = {\nu_{\rm obs}\over c^{3}\rho_{\rm c}}F_{\nu} =
{4\pi^{2}\over
3H^{2}}\nu_{\rm obs}^{2} h_{\rm BH}^{2}.
\end{equation}

\noindent Assuming a maximum efficiency for the generation of GWs
($\varepsilon \simeq 7\times 10^{-4}$; Stark \& Piran 1986) during
the collapse of an object to a black hole, one has $\Omega_{\rm
GW}<10^{-6}$ for the redshifts and masses studied here.

In Fig. 2 we present the amplitude $h_{\rm BH}$ as a function of
the observed frequency ($\nu_{\rm obs}$) for different values of
$\varepsilon$, SMBH mass and formation redshift. We also present
the {\it LISA} sensitivity  ($h_{\rm s}$) for a signal-to-noise
ratio of 1 for burst sources.

For example, $h_{\rm BH}> h_{\rm s}$ for $M_{\rm BH}=10^6
M_\odot$ and $\varepsilon > 10^{-5}$. Thus, even for low GW
efficiency the signal produced by these SMBHs could be detected by
{\it LISA}.

\section{Discussion and Conclusions}

The results presented here were obtained for an open Universe
model with $\Omega_{\rm b} = 0.1$ and $H_0 = 50\,{\rm km}\;{\rm
s}^{-1} {\rm Mpc}^{-1}$. We also assume the same scenario as FS,
with all the SMBHs forming at the same redshift. For a given event
rate, and for a given range of mass, we first calculate the
duty-cycle to see whether the GWs produced by the ensemble of
SMBHs generate a stochastic background. For an event rate
exceeding $1-10$ day$^{-1}$ we find that the bursts do not overlap
and so they do not produce a continuous stochastic background.  In
particular, a stochastic background could occur for black holes
with $M_{\rm BH} \sim 10^7 M_\odot$ only if the event rate
exceeded  30 per day$^{-1}$. In this case we would have
$\tau_0/\Delta t_0 > 1$ and the GWs of different seeds could
overlap producing a background with amplitude given by equation
(7). SMBHs formed with masses $< 10^6M_\odot$ could produce a GW
background for the same event rate only if they formed at $z>5$.

The major advantage of the FS scenario, as a cosmological source
of $\gamma$-ray emission, is its enormous energy reservoir; the
gravitational binding energy is $E_{\rm g}\sim 10^{54}(M_{\rm
BH}/M_\odot)\;{\rm erg}$. Another advantage of this scenario is
related to the angular scale of the sources. Although tremendous
energy is deposited into the fireball ($\sim 10^{52}\,\,{\rm
ergs}$ during the collapse to a black hole of $10^6M_\odot$), the
distortion produced in the cosmic background radiation through the
scattering of hot electrons (Sunyaev-Zeldovich effect) occurs on a
very small angular scale ($\theta \; \lesssim \; 10^{-10}$ arc
seconds) and is therefore undetectable.

In the FS model a potential problem, as a GRB source, is related
to the ``baryon-loading" \footnote{There are many papers in the
literature discussing aspects related to the injection of energy
(including the baryon-loading problem) associated with GRBs. In
particular, we refer the reader for the papers of Shemi \& Piran
1990; Kobayashi, Piran \& Sari 1999 and Fuller, Pruet \& Abazajian
2000.}, that is, the confinement of the electron/positron/photon
fireball by the baryons which could carry energy of it in the form
of kinetic energy, thus diminishing the amount of energetic
photons (the gamma ones). This suggests that the region at several
Schwarzschild radii from the supermassive star core should have
extremely low baryon density. There are, at least two ways to
avoid the excessive baryon-loading: rotation of the star producing
the flattened collapse or the collapse of a dense star cluster
instead of a single object. This could result in a different event
rate for the GRBs and the GW bursts, not all GW bursts being
related to GRBs in the present scenario since the baryons could
block the $\gamma-$ray.

Even if the GRBs and GW bursts have completely different event
rates, either because the source of GWs does not produce GRB at
all or because the gamma radiation is blocked, it would be
possible to verify the FS scenario by looking for GRBs once GW
bursts associated with SMBH formation are observed and identified.
There will be a time interval between the GRB and the GW burst
because the types of radiation are generated in different ways.
The generation of the GRB depends on a series of physical
processes after the collapse of the core, e.g., the generation of
the fireball to accelerate the matter to the ultra-relativistic
regime when the kinetic energy in the fireball could be converted
to $\gamma$-rays. The GWs, on the other hand, are mainly produced
when the SMBH is formed, through the excitation of its
quasi-normal modes. A detailed modeling is required however to
evaluate the time interval between the GRB and the GW burst.

Using the {\it LISA} observatory to detect GW  bursts related to
the SMBHs formation, one could find their GW amplitudes, the
characteristic frequencies and also the formation rate of SMBHs.
If we also find the redshift associated with the events (by
observing in the electromagnetic window) we will be able to obtain
the SMBH masses and the GW efficiency using the model proposed
here. By comparing the SMBH formation GW event rates with the GRB
rates one could also infer what fraction of an ensemble of SMBHs
had conditions to generate GRBs and to impose constraints on the
FS scenario. Then in the present study we are proposing an
independent way to check FS model through GW astronomy.

\acknowledgements The Brazilian agency FAPESP partially supports
this work (JCNA under grants 97/06024-4 and 97/13720-7; ODM under
grant 98/13735-7). ODA thanks CNPq (Brazil) for financial support
(grant 300619/92-8). We would like to thank Dr. Robin Stebbins and
Prof. Peter Bender for kindly providing us with the {\it LISA}
sensitivity curve. Finally, we would like to thank an anonymous
referee for his (her) criticisms and suggestions which greatly
improved the present version of our paper.

\clearpage

\begin{figure}
\begin{center}
\leavevmode
\centerline{\epsfig{figure=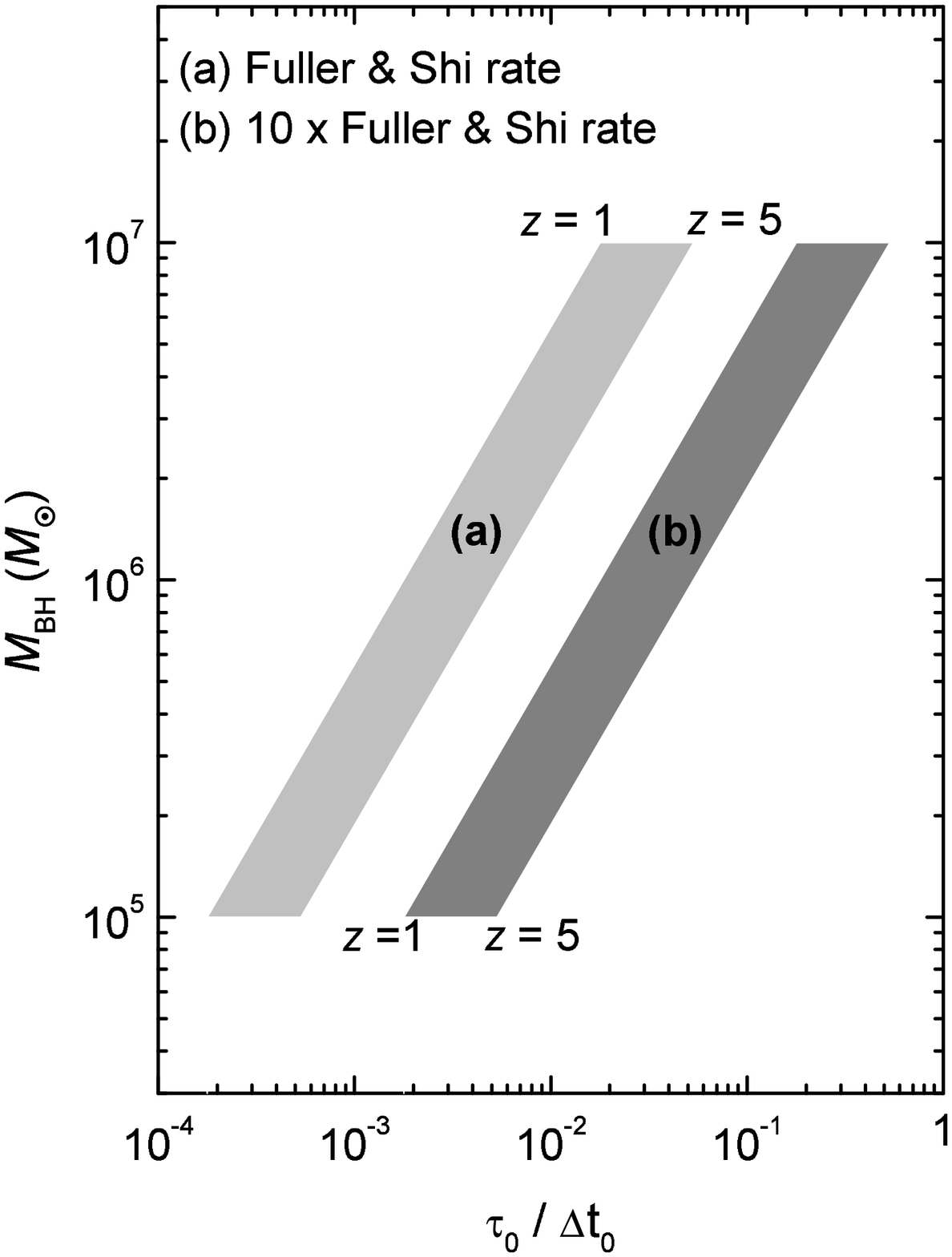,angle=360,height=20cm,width=14cm}}
\caption{Duty-cycle versus the mass of SMBHs for the formation
redshift range $z=1-5$. The results are presented for 1 and 10
events day$^{-1}$. The cosmological model considered has $\Omega_0
= \Omega_{\rm b}=0.1$ and $H_0 = 50\,{\rm km}\;{\rm s}^{-1} {\rm
Mpc}^{-1}$.}
\end{center}
\end{figure}

\clearpage

\begin{figure}
\begin{center}
\leavevmode
\centerline{\epsfig{figure=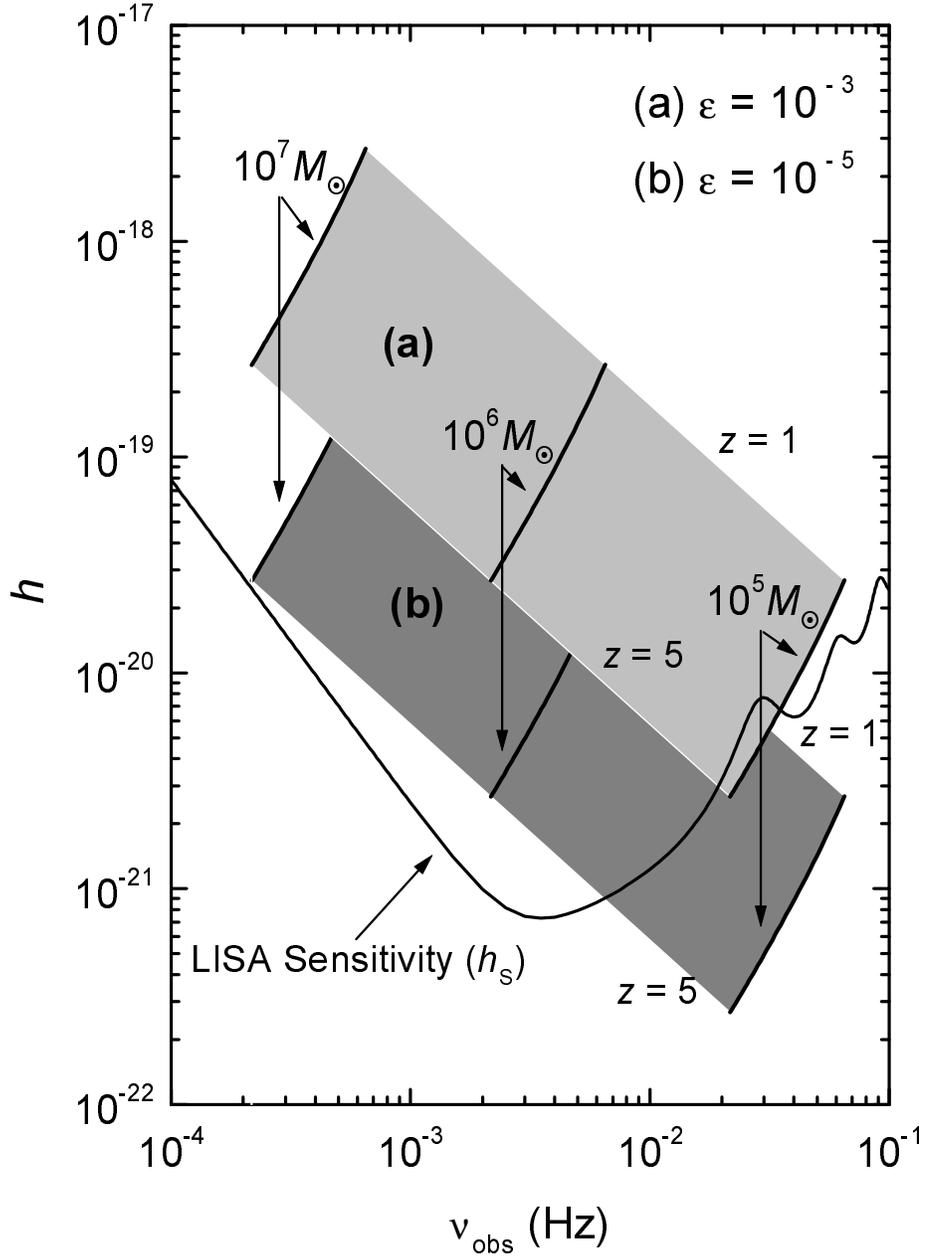,angle=360,height=20cm,width=14cm}}
\caption{Dimensionless amplitude $h_{\rm BH}$ as a function of
observed frequency for $\varepsilon = 10^{-5}$  and $\varepsilon =
10^{-3}$ for the burst of GWs for $M_{\rm BH}=10^5, \, 10^6\, {\rm
and}\, 10^7 M_\odot$ at redshifts $z = 1 - 5$. The {\it LISA}
sensitivity for burst sources ($h_{\rm S}$) is also plotted. The
cosmological model considered has $\Omega_0 = \Omega_{\rm b}=0.1$
and $H_0 = 50\,{\rm km}\;{\rm s}^{-1} {\rm Mpc}^{-1}$.}
\end{center}
\end{figure}

\end{document}